# Internet of Things for Residential Areas: Toward Personalized Energy Management Using Big Data

Hojjat Salehinejad

Farnoud Data Communications Company (FADCOM), Kerman, Iran

**Abstract**

Intelligent management of machines, particularly in a residence area, has been of interest for many years. However, such system design has always been limited to simple control of machines from a local area or remotely from the Internet. In this report, for the first time, an intelligent system is proposed, where not only provides intelligent control ability of machines to user, but also utilizes big data and optimization techniques to provide promotional offers to the user to optimize energy consumption of machines. Since a high traffic communication is involved among the machines and the optimization-big data core of system, the communication core of the proposed system is designed based on cloud, where many challenging issues such as spectrum assignment and resource management are involved. To deal with that, the communication network in the home area network (HAN) is designed based on the cognitive radio system, where a new spectrum assignment method based on the ant colony optimization (ACO) algorithm is proposed to perform spectrum assignment to the machines in the HAN. Performance evaluation of the proposed spectrum assignment method shows its performance in fair spectrum assignment among machines.

## I. Introduction

The recent advancements in information and communication technology (ICT) and dramatically increment in using ICT based services such as social networks have caused rapid generation, transmission, processing, and storage of data around the world. Meanwhile, cities are growing faster than before and it is estimated that by 2050 two-thirds of people will be city-dwellers [16]. Cities require rapid technological development, driven by cloud-based services and more powerful mobile devices, big data and analytics. In this emerging "Networked Society", continuous development of ICT infrastructures, such as data centres are required to provide fast and reliable services for the users [17]. Intelligent networks will be critical to the basic functioning of cities and to their success in meeting current and emerging challenges. More technology development requires more energy consumption and as a result, more greenhouse gases (GhG) emissions [18]. The GhG emission and particularly carbon emissions are major contributors to the global warming issue. The technology has become a part of everyone's daily life and it is not possible to stop or limit its development, particularly the data centres and ICT technology. This is while there is no far and safe place in the earth to release the wastes without negative impacts on environment.

In "Networked Society", a part of the main attributes is maintaining sustainable economic, social and environmental developments. In such society, cities must provide high quality services in different areas such as governmental, healthcare, educational and other public services. This quality improvement must be conducted with keeping the environment safe from emissions, climate change, and industrial footprints [17]. Unfortunately, ICT is one of the main reasons for climate change [18]. However, optimal green design of ICT infrastructures in "Networked Society" can meet challenges encountered to cities. It can improve efficiency in the delivery of services and productivity, as well as encourage new collaboration and innovation that fuel socio-economic development [17].

The ICT industry is one of the major contributors in attempts against climate change. It is estimated that the ICT can reduce emission of 7.8 Gt CO2e in 2020, which is approximately $946.5 billion energy efficiency [20]. The ICT can contribute in tackling the carbon reduction based on the foundations of "Sustainable Networked Society". The ICT can be utilized in "Sustainable Networked Society" in a smart way to form the future smart cities. The smart ICT can provide smart approaches in a variety of services and applications such as climate, energy, transportation, public sector, health, education, sustainable lifestyle, safety, and security [19]. The renewable portfolio standards (RPS) mechanism generally places an obligation on electricity supply companies to provide a minimum percentage of their electricity from approved renewable energy sources [22]. According to the U.S. Environmental Protection Agency [23], as of August 2008, 32 states plus the District of Columbia had established RPS targets. The RPS targets currently range from a low of 2% to a high of 25% of electricity generation, with California leading the pact that requires 20% of the energy supply come from renewable resources by 2010 and 33% by 2020. The RPS non-compliance penalties imposed by states range from $10 to $55 per megawatt-hour [23]. In order to develop green energy in distributed manners, mostly for smart city applications, new topologies and policies must be developed. The modernized electricity generation, transmission, and distribution based on the ICT, the smart grids [21], can be considered as one of the main parts of smart cities for the climate and energy schemes. The smart grids in "Sustainable Networked Society" can provide smart energy solutions that are more efficient and can improve currently very inefficient delivery in a fossil fuel-based economy. The energy and water resources can be gathered, tracked, controlled, and reconfigured optimally based on information management systems. By developing smart ICT-based topologies for integrated transportation and communication in "Sustainable Networked Society", cities can connect to each other as well as suburbs. One step forward, the paper usage and travels can be reduced dramatically by integration of E-government services. The travels in smart cities can even be more reduced by developing digital health and remote monitoring solutions, such as telemedicine schemes [24]. To develop health in "Sustainable Networked Society", the daily food initiatives can be initiated based on low-carbon/water footprint products to not only provide health food, but also avoid waste and contaminating environment [22].

To develop "Sustainable Networked Society", it is necessary to create frameworks that makes the link between ICT policies and policies in other areas such as education, health, and transportation. The smart ICT infrastructures also need to be designed to be integrated into the existing systems conveniently [19]. To do so, many challenges exist such as:

- Resource management in integration of smart ICT infrastructure into different collaborators such as the health, transportation, and enterprises.
- Optimal planning of smart ICT infrastructures, meeting future requirements.
- Introduction of new business models for ICT providers.
- Providing security for smart ICT infrastructure deployment.
- Optimal design of eco-systems and services in large-scale.
- Categorizing end-users and devices in the smart city for specific applications.
- Development of smart grids on the "Sustainable Networked Society" platform.

In order to address solutions for challenges facing the development of "Sustainable Networked Society", many well-developed approaches can be utilized or new schemes can be developed such as in machine learning, optimization, and data mining. The nature of the "Sustainable Networked Society" is for large-scale developments with many integrated parameters and objective. In order to achieve reasonable optimized designs, with respect to all possible constraints, specific methodologies are required. Evolutionary algorithms are promising tool for solving multi-dimensional, large-scale, non-linear, non-convex, combinatorial optimization problems [25], [26], [28]. Therefore, such optimization tools will be considered and developed in this research to design the "Sustainable Networked Society".

The rest of the paper is organized as follows. In section II, a review on related works in the field are presented. In section III, the proposed architecture for the machine to machine communication on cloud is presented. As a case study, the channel assignment problem in machine to machine communication systems in home area networks (HAN) based on the ant colony system (ACS) is studied. To do so, a brief survey on ACS is presented in Section IV and the cognitive system allocation model for HAN is presented. In section V, the proposed ACS spectrum allocation method, as a part of the proposed architecture, is discussed in details and its performance is evaluated for varying number of users and channel availability scenarios. Finally, some interesting forthcoming research challenges are introduced and the paper is concluded in section VI.

## II. Related Works

Many topologies have been proposed for machine to machine communications; however, none of them have considered a complete structure for energy efficiency and control in the machines through cloud communications. As the state-of-the-art research works show, one of the most important parts in such topologies is the communications system between machines. In communications, sensing, transmission, and control are introduced as the main functionalities [1]. Since in communication systems, such factors are deployed in a large geographical field with a large number of nodes, the communications infrastructure has to integrate enabling networking technologies to cover the entire region and fulfill expectation of secure and reliable communications [3]. However, due to the unique challenges imposed on the communication systems such as limited available spectrum, inefficiency in spectrum usage, enormous amount of data, highly varying traffic, quality of service (QoS) issues, and etc., the existing communications network is infeasible and cannot be applied trivially. Such challenges necessitate a new communication paradigm, referred to as cognitive radio networks and dynamic spectrum access (DSA) to exploit the existing wireless spectrum opportunistically [5,6]. The DSA stands for the opposite of the current static spectrum management policy which allows the cognitive radio to operate in the best available channel [7]. The cognitive radio is a context-aware system which is capable of reconfiguration based on the surrounding environments and their own properties [8]. Generally, this technology is divided into four stages which are spectrum sensing, spectrum management, spectrum sharing, and spectrum mobility [5,6]. In [1], a revolutionary communication architecture based on cognitive radio is introduced for efficient, sustainable, secure, and stable communications. In this architecture, the entire smart grid communication architecture is divided into three cognitive radio based hierarchical layers which are home area networks (HANs), neighborhood area networks (NANs), and wide area networks (WAN) [1,3]. Generally, a HAN is comprised of intelligent devices such as a cognitive home gateway (HGW), smart meters, and sensors. The HGW is the cognitive data access point responsible for collecting the HAN data such as energy consumption information and transmitting it toward the NAN. The NAN cognitive gateway (NGW) then collects several HANs data and delivers them via WANs to a control center. In this architecture, the NGW is considered as the cognitive radio access point to provide single-hop connection with HGWs in a hybrid access manner and distributes spectrum bands to them.

In cognitive radio system principles, two kind of users, which are the primary users (PUs) and secondary users (SUs), exist. The PU refers to licensed user of any legacy licensed spectrum system such as TV or telecommunication operators. This user has exclusive opportunity to access the assigned spectrum. The SU refers to the unlicensed user, which can only opportunistically access the spectrum holes that are not used by the PUs. In a machine to machine communication environment, due to the large number of users and lack of spectrum availability, licensed spectrum bands are not enough to meet the large amount of data transfer. Therefore, unlicensed access is also needed for the HGWs to improve the capacity and throughput of the NAN. In unlicensed access, the HGWs and NGWs could be considered as SUs, and the unoccupied spectrum bands are assigned to them as communication channels in an opportunistic manner [1]. In [1,3], a hybrid dynamic spectrum access (HDSA) methodology in NANs is proposed to significantly improve the flexibility of communications infrastructure and spectrum efficiency. In this structure, the NGW is responsible for allocating the spectrum bands to the HGWs within its area. The HGWs being allocated with unlicensed bands act as SUs. An in- service SU has to hand off to a spectrum hole once a PU appears and occupies its spectrum band. If there is no spectrum hole available to which to hand off to, the SU will be dropped. Due to the dynamic nature of spectrum availability, which causes a serious difficulty in stable and assured QoS provisioning, a hybrid guard channel (HGC) strategy is proposed in [1] to protect the current services and sustain their QoS at a satisfactory level. In the traditional guard channel (GC) scheme, a number of channels are reserved for handoff traffic and new services are not allowed to use the reserved channels. The HGC scheme has added the aptitude of reserving a number of channels for the usage of spectrum handoff for both the licensed and unlicensed bands. Based on this strategy, four types of channels in HDSA is hosted which are licensed guard channels (GCs), unlicensed GCs, licensed common channels, and unlicensed common channels. The challenge that arises here is fair spectrum resource management among HGWs in the NAN level and NGWs in the WAN level. However, isolated spectrum assignment for WAN and NAN may reduce the extensive system performance, since most of the NANs located in a WAN have access to similar spectrum resources. This results in a competition between NANs on available resources. The importance of this issue arises as all the NANs in a WAN cover diverse number of HANs and therefore their traffic flow and demand for spectrum bands are different. Therefore, the concept of joint WAN/NAN spectrum management is introduced in [1] at the WAN level. In this strategy, there is a spectrum broker server to manage the spectrum resources of the entire communications infrastructure of the smart grid. From the perspective of the utility companies, the communications infrastructure should operate economically, efficiently, and adaptively. The spectrum management of the WANs and NANs should be jointly optimized. Due to the enormous geographical scale of the problem and large number of users in the topology, the above problem can be considered as a NP-hard optimization problem, which requires considering all possible resource management states. Similar situation can occurs for wireless sensor networks (WSNs) [29].

Metaheuristic optimization algorithms are of great interest for solving large scale problems and particularly the NP-hard ones in academia and industry [9-11]. The crucial benefits of such nondeterministic structures are their less complexity, low computational load, and time usage for solving complex problems, which may have no derivatives, in a short period [9-11]. In [11] a spectrum allocation model is initially presented, and then spectrum allocation methods based on genetic algorithm (GA), quantum genetic algorithm (QGA), and particle swarm optimization (PSO), are proposed. The D-ring method is employed in [12] for channel assignment in wireless mobile networks. The ant colony system (ACS) is used for spectrum allocation in cognitive networks in [6]. In [13], the spectrum assignment problem is mapped to a graph coloring problem (GCP) and the ACS strategy is employed for solving it in [14] for mobile cognitive networks. By considering the cognitive radio based hierarchical communications architecture in [1], in this work the graph theory is employed for mapping the HAN topology into a graph problem. Then a top-down joint fair HAN cognitive resource management procedure based on the ACS strategy is proposed.

### III. Proposed Architecture

### 3.1. Architecture Description

In this section, the proposed intelligent residence topology is presented. As it is demonstrated in Figure **1**, the machines {$M_{N,1},…,M_{N,P1}$} in residence $N$ are connected directly to the specific gateway $Gateway_N$. The user $N$ has direct access to the $Gateway_N$ for controlling and communication with other parts of the network. In order to present a design on top of the internet available at residence via $Router_N$ and also provide security for the home area network (HAN) where the machines {$M_{N,1},…,M_{N,P1}$} are located, the $Gateway_N$ is the only pathway between the HAN and outside world. The all the routers are connected through a cloud network to the data center, where the data of machines are recorded. Since the collected data from machines have a huge size, where a portion of that is redundant data, the data processing approaches are employed to reduce dimensionality of data and provide clean data to the optimizer. The optimizer, based on the received patterns of each machine usage, the user behavior, and recommended policies, provides promotional plans to the user and the machines.

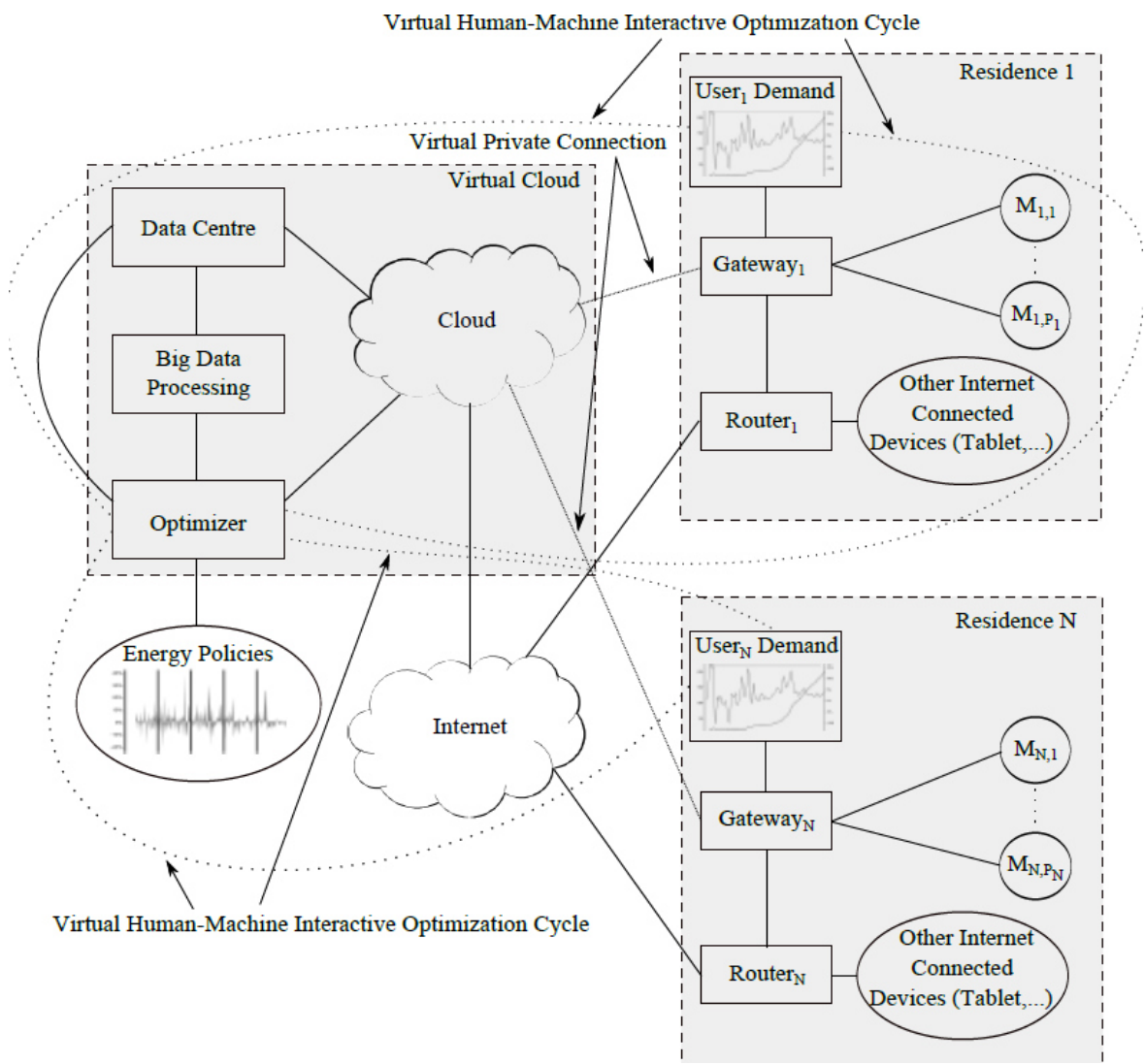

Figure 1 Proposed Intelligent Residence Topology.

In this topology, the data center and optimizer are in direct communication with each other,

with provides a feedback from users to the optimizer. In Figure 2, the message exchanging procedure between different components of the system is presented. The working hours and device usage of devices are mostly managed by users, for example when to turn on the washing machine or ironing the cloths. The energy consumption as well as usage time of devices are sent as raw data to the Gateway controller, which is in charge of receiving the machine data, collecting data from all the machines in the HAN, forwarding collected data toward the data center through the router, and sending controlling commands to the machines. The router, is considered as a regular one, which not only routes the data from/to Gateway, but also routes the other traffic (e.g. Internet usage). The router sends the data through the cloud to the data center for storage and further processing. The collected data from users and machines in the network is high dimensional, noisy and has a huge size. Therefore, big data processing techniques employed by the processing unit to clean the data (e.g. redundancy) and prepare the data for optimization stage. Based on the recommendation from specific sources such as the Hydro companies, the optimization stage tries to optimize energy consumption of machines and make recommendations to the users. The user has the authority to accept, decline, or modify the recommended plan. After approval, the commands are forwarded toward the target machine for operation and an acknowledgment is sent to the user.

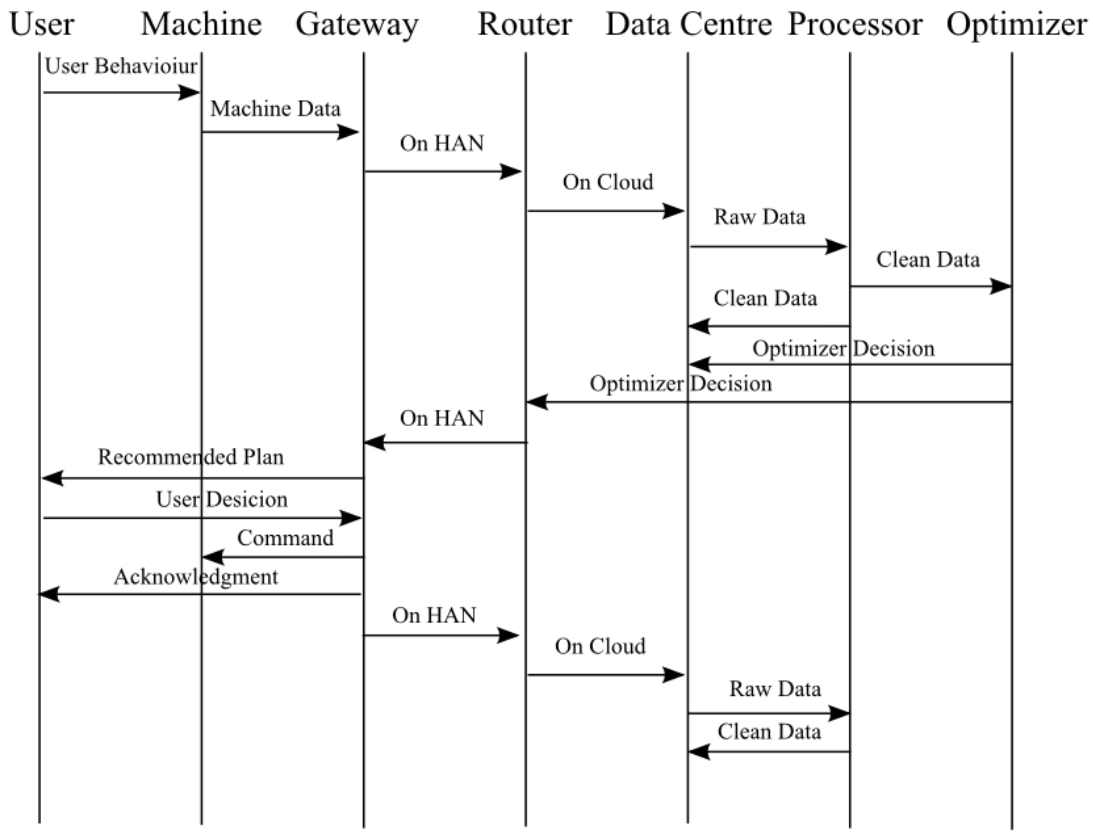

*Figure 2 Diagram of the message exchanges between system components.*

## 3.2. Challenges

As it is demonstrated, many challenges are facing the proposed architecture which are stated as below:

3.2.1. Optimizer

Many optimization algorithms have been proposed for different applications in the literature. The question which arises here is what optimization algorithm is suitable for each application. There are many parameters that specify the optimization problem, and the art of optimization is in selecting the best approach which matches the criteria of the problem. The optimization problems are specified by number of dimensions, linearity/non-linearity, modality, convex/non-convex, and constraints condition, to mention some. In the proposed topology, we are dealing with high dimensional, large-scale, non-linear, non-convex optimization problems, which can be classified as NP-hard problem. From other, perspective, the computational complexity of the algorithm is of great importance, since the core system receives a huge amount of data, which need to be analyzed and optimized. Therefore, parallel processing method on could can be utilized to deal with the problem. Metaheuristic algorithms, such as evolutionary algorithms, are state of the art method to deal with such problems. Particularly that in population based algorithm, we are dealing with a set of population, where the population set can be divided to different groups of agents and run on separate cluster on the could for parallelization.

3.2.2. Data Centre

The amount of data as well as its rapid transmission from machine toward the data center are big challenges for data centers. The system have to be able to iterate on functionality quickly. It should be developed in a much more agile manner, with dynamic reallocation capability. The huge volume, dimensionality, and complexity of data under management is of great importance. Although the scale of data is important, but complexity of data is a serious problem. Today's data is consisted of a mix of structured and unstructured components. Most database technologies are not architected to allow elastic expansion or contraction of capacity or compute power, which makes it hard to achieve many of the benefits (and cost savings) of cloud technologies [27].

3.2.4. Resource Allocation in Cloud

In the new machine to machine communication paradigm, there are many devices that need to be supported through communication infrastructure. The limited available spectrum and the inefficiency in spectrum usage necessitate a new communication paradigm, referred to cognitive radio networks as well as next generation (XG) networks and dynamic spectrum access (DSA), to exploit the existing wireless spectrum opportunistically [1]. Cognitive radio techniques provide capability to use or share spectrums in an opportunistic manner. The DSA stands for the opposite of the current static spectrum management policy [2], [4]. The DSA techniques allow the cognitive radio to operate in the best available channel. Generally, the cognitive radio technology is consisted of four stages which are spectrum sensing, spectrum management, spectrum sharing, and spectrum mobility. Open Spectrum allows unlicensed (secondary) users to share spectrum with legacy (primary) spectrum users. Based on agreements and constraints imposed by primary users, secondary users opportunistically utilize unused licensed spectrum on a non-interfering or leasing basis at any location over the entire spectrum. While maximizing spectrum utilization is the primary goal of open spectrum systems, a good allocation scheme also needs to provide fairness across devices.

## IV. Channel Allocation In The Proposed Machine To Machine Architecture: A Case Study

In this section, before detailed discussion of the proposed algorithm, a brief survey on ACS is presented. Then, the cognitive channel allocation model for HAN is presented. The proposed algorithm to solve the model is presented at the end.

### 4.1. Ant Colony System

The ant colony optimization (ACO) is a class of algorithm whose first member, called ant system (AS), was initially proposed by Colorni, Dorigo and Maniezzo [15]. Although real ants are blind, they are capable of finding shortest path from their nest to a food source by exploiting information of a liquid substance, called pheromone, which they release on the transit route. The more developed AS strategy attempts to simulate the behavior of real ants with the addition of several artificial characteristics: visibility, memory and discrete time to resolve many complex problems successfully such as the traveling salesman problem (TSP) [15] and transportation networks [16]. Even though many changes have been made to the ACO algorithms during the past years, their fundamental ant behavioral mechanism that is a positive feedback process demonstrated by a colony of ants, is still the same. Ants algorithm finds plenty of applications in different areas of wireless communications such as routing and resource management [17,18], cell planning and user detection [19, 20] and spectrum assignment [13, 14]. Different steps of a simple ACS algorithm are as follow:

*Problem Graph Depiction:* Artificial ants move between discrete states in discrete environments. Since the problems solved by ACS algorithm are often discrete, they can be represented by a graph with $N$ nodes and $R$ routes.

*Ants Distribution Initializing:* A number of ants are placed on the origin nodes. The number of ants is often determined by trial and error and number of nodes in the region.

*Ants Probability Distribution Rule:* Ants probabilistic transition between nodes can also be specified as node transition rule. The transition probability of ant $k$ from node $i$ to node $j$ is given by

$$p_{ij}^k = \begin{cases} \dfrac{(\tau_{ij})^\alpha (\eta_{ij})^\beta}{\sum_{h \notin tabu_k} (\tau_{ih})^\alpha (\eta_{ih})^\beta} & j \notin tabu_k \\ 0 & otherwise \end{cases} \qquad (1)$$

where $\tau_{ij}$ and $\eta_{ij}$ are the pheromone intensity and the heuristic visibility (cost) of direct route between nodes $i$ and $j$ respectively. Relative importance of $\tau_{ij}$ and $\eta_{ij}$ are controlled by parameters $\alpha$ and $\beta$ respectively. The $tabu_k$ is a set of unavailable routes for ant $k$.

*Update Global Trail:* When every ant has assembled a solution at the end of each cycle, the intensity of pheromone is updated by a pheromone trail updating rule. This rule for ACS algorithm is given as

$$\tau_{ij}^{new} = (1-\rho)\tau_{ij}^{old} + \sum_{k=1}^{m} \Delta\tau_{ij}^k \qquad (2)$$

where $0 < \rho < 1$ is a constant parameter named pheromone evaporation. The amount of pheromone laid on the route between nodes $i$ and $j$ by ant $k$ is computed from

$$\Delta\tau_{ij}^k = \begin{cases} \dfrac{Q}{f_k} & \text{if route } (i,j) \text{ is traversed by the } k^{th} \text{ ant (at the current cycle)} \\ 0 & otherwise \end{cases} \qquad (3)$$

where $Q$ is a constant parameter and $f_k$ is the cost value of the found solution by ant $k$.

*Stopping Procedure:* This procedure is completed after arriving at a predefined number of cycles, or the maximum number of cycles between two improvements of the global best solutions.

## 4.2. Cognitive HAN Resource Allocation Model

In a typical HAN, the HGW manages the spectrum resources and allocates channel to smart machines in a centralized mode as in Figure 3. In such scenario, the smart devices interfere with each other on available resources. This interference can be modeled as a graph problem as follow.

### 4.2.1. HAN as a Graph Problem

The main idea of efficient utilization in an open spectrum is to find a suitable assignment of channels to the SUs while minimizing the interference among them. In traditional spectrum assignment procedure [14,15], when two simultaneous transmissions overlap in spectrum and physical location, both can fail. Hence, a user seizing spectrum without coordinating with others can cause harmful interference with its neighbors and degrade overall spectrum usage. To have a closer look at a sample scenario, some PUs and SUs (i.e. smart devices) are located randomly in a HAN as illustrated in Figure 4. For the sake of simplicity, the users are considered static, while in real scenarios mobile users such as hybrid cars can also be considered as smart grid users which will be considered in further works. In this scenario, two PUs exist where each one has occupied one licensed channel from the spectrum pool and cannot use the channel taken by the other one. The SUs are assumed to use orthogonal frequency division multiple access (OFDMA) technology. The smart grid SUs are waiting for channel within the coverage area of PUs. Each $PU_x$ inhabits a channel $m$ with a radius of protection area $dp_{x,m}$. The radius of coverage area for SU $n$ on channel $m$ is defined by $ds_{n,m}$. This radius is adjustable by tuning the transmit power on channel $m$ and is defined as

$$ds_{n,m} = \min(d_{\max}, \min(Dist_{x,n} - dp_{x,m})) \qquad (4)$$

where $Dist_{x,n}$ is the distance between PU $x$ and SU $n$. As a general rule, the interference range is bounded by the minimum and maximum transmit power that is $[d_{\min}, d_{\max}]$, [15]. Increasing coverage area of a SU results in interference probability with a neighboring smart device user. For simplicity, SUs are assumed with a fixed power control scheme to adjust their transmit power to the maximum permissible level in order to avoid interfering with PUs. In addition, the transmission and interference ranges are assumed identical. Any radiation from each PU or SU into the coverage area of other user on channel $m$ would cause interference, [15]. Typically, a HAN is consisted of totally $M$ channels. In the binary channel availability matrix of users $L_{N \times M}$, if $l_{n,m} = 1$ channel $m$ is available for $SU_n$. The interferences between different SUs on channels are determined by the binary matrix $C_{N \times N \times M}$ where if $c_{n,k,m} = 1$ SUs $n$ and $k$ interfere on channel $m$. By considering the interferences, the topology in Figure 4 can be modeled as a graph problem where each SU denotes a node with a pool of available channels and each interference between them on a specific channel as an edge. In a GCP [15], each vertex is colored using a number of colors from its available color list under the constraint that two vertices linked by an edge cannot share the same color. In GCP, the objective is to assign a color to each SU so that a given utility function maximizes [11]. Therefore, the spectrum allocation problem is mapped into a graph problem defined with $G = (V, L, C)$, where $V$ is a set of nodes denoting the SUs, $L$ is list of available channels, and $C$ is interference matrix between SUs.

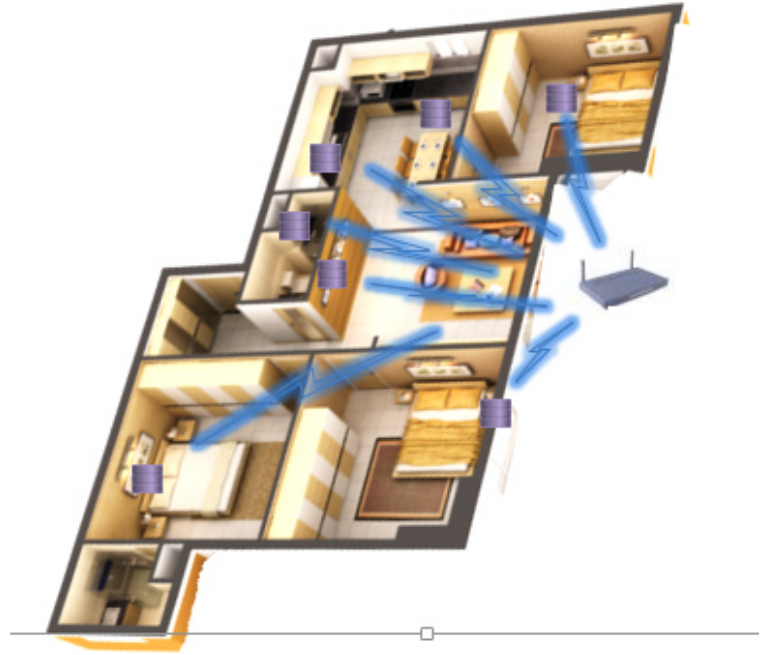

*Figure 3 Home area network (HAN) architecture.*

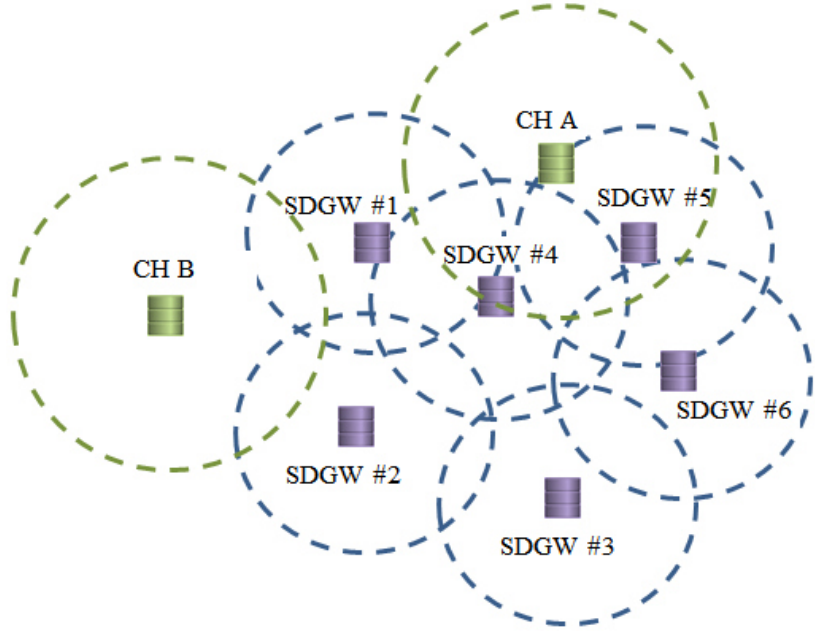

Figure 4 Interferences between smart device gate ways (SDGWs) in a home area network (HAN).

### 4.2.2. Network Spectrum Utilization Measure

In a spectrum allocation problem, the final channel assignment is represented by a binary matrix $A_{N \times M}$. The binary $a_{n,m} = 1$ means channel $m$ is allocated to SU $n$. In some situations where no channel is assigned to a user, it is called a starved user. A reward vector is defined for each SU when obtaining a channel $m$ as $R = \{r_n = \sum_{m=1}^{M} a_{n,m} b_{n,m}\}_{N \times 1}$ where $B = \{b_{n,m}\}_{N \times M}$ can be considered as maximum bandwidth or throughput that can be acquired (assuming no interference from neighbors) by user $n$ using channel $m$. Therefore, the reward can be the capacity of using a channel (assuming the signal to noise ratio (SNR) is a function of $ds_{n,m}$) defined as

$$b_{n,m} = \log(1 + f(ds_{n,m})), d_{min} \leq ds_{n,m} \leq d_{max} \tag{5}$$

or can be considered as coverage of SU $n$ using channel $m$ as

$$b_{n,m} = \begin{cases} (ds_{n,m})^2 & d_{\min} \leq ds_{n,m} \leq d_{\max} \\ 0 & l_{n,m} = 0 \end{cases}. \quad (6)$$

By defending the rewards, the spectrum allocation problem can be presented as the optimization problem

$$A^* = \arg\max_{\substack{A \in \Lambda_{L,C} \\ i \in \{MSR, MMR, MPF\}}} U_i(R) \quad (7)$$

where $\Lambda_{L,C}$ is set of conflict free channel assignment for a given $L$ and $C$. The $U_i(R)$ is defined as the network utilization where three measures called max-sum-reward (MSR), max-min-reward (MMR), and max-proportional-fair (MPF) are considered [9, 15]. The MSR function maximizes the total spectrum utilization in the system regardless of fairness as

$$U_{MSR}(R) = \sum_{n=1}^{N} r_n. \quad (8)$$

The MMR function maximizes the spectrum utilization regarding the user with the least allotted spectrum as

$$U_{MMR}(R) = \min_{1 \leq n \leq N}(r_n). \quad (9)$$

This function gives the user with the lowest reward, the largest possible share while not wasting any network resources. The MPF function addresses fairness for single-hop flows as

$$U_{MPF}(R) = (\prod_{n=1}^{N}(r_n + 10^{-6}))^{\frac{1}{N}}. \quad (10)$$

Proofs and more details are available in [15].

### 4.3. Cognitive HAN Fair Resource Management Algorithm

By modeling the resource management problem as a graph problem at the NAN level, the total structure of a joint WAN/NAN can be modeled as in Fig. 4. In order to assign the channels fairly and efficiently, the algorithm in Fig. 5 is proposed. In this ACS procedure, the spectrum breaker is considered as the artificial ants nest and the nodes in NANs are considered as the food sources. Therefore, the ants are run from the spectrum breaker, each carrying the candidate channel and move toward the most possible HGW based on the motivation defined. The detailed steps of the algorithm are described as follows.

*Initialization:* As the first step, initial value of the parameters such as pheromone matrix

$T_{N_{HGW} \times M \times N_{NGW}} = 1$, selection parameters $G$ and $G'$, and the evaporation coefficient $\rho$ are set.

*Graph Mapping:* At the beginning of each procedure, topology information is gathered and the interferences among the SUs is mapped into a graph problem, as described in the first part of section II.

*NGWs Probabilities:* As it is illustrated in Fig. 4, in each iteration an ant, carrying a channel, is released from the spectrum breaker toward a HGW. At the first stage each ant must select a NGW. Therefore, it calculates probability of each candidate NGW as

$$P_{NGW}^i \bigg|_{\forall i \in \Gamma_{1 \times N_{NGW}}} = \begin{cases} \dfrac{\sum_{j=1}^{N_{HGW}^i} s_j^i p_{j,m}^i}{\sum_{i=1}^{N_{NGW}} \sum_{j=1}^{N_{HGW}^i} s_j^i p_{j,m}^i} & N_{HGW}^i \geq 1 \\ 0 & N_{HGW}^i = 0 \end{cases} \quad (11)$$

where $\Gamma_{1 \times N_{NGW}}$ is list of available NANs, $N_{NGW}$ is number of NANs, and $N_{HGW}$ is number of HGWs in NAN $i$. The probability of NGW $j$ for channel $m$ is defined by

$$p_{j,m}^i = \dfrac{T_{j,m,\xi}^i B_{j,m}^i}{M N_{HGW}^i B_{\max}} \sum_{l=1}^{M} L_{j,l} \sum_{k=1}^{N_{HGW}^i} C_{j,k,m} \quad (12)$$

where $\xi$ is the iteration number.

Based on the operations of HGC as described in [1,3], a status parameter $s_j^i$ is developed for each HGW $j$ in NAN $i$. With respect to the HGC policy, two smart grid service (SGS) which are new SGS (NSGS) and handoff SGS (HSGS) are defined. If the statuses in Table I are satisfied for a SGS on channel $m$, $s_j^i = 1$ and the SGS is permitted to access the network. Otherwise, $s_j^i = 0$ and the SGS $j$ access is blocked.

*NGW Selection:* The random parameter $g$ with uniform probability in $0 < g < 1$ is compared with the selection parameter $G$ with uniform probability in $0 < G < 1$. The result picks up one of the following two selection methods

$$S_{NGW} = \begin{cases} \arg\max \left( P_{NGW}^{i} \right) & g > G \\ \text{Roulette Wheel}\left( P_{NGW}^{i} \right) & \text{otherwise} \end{cases} \quad (13)$$

where $S_{NGW}$ is the selected NGW by the ant.

*Semi-Local Pheromone Updating:* After selection of a NGW, with respect to the ACS policy as in [6,9,14], pheromone of all HANs in the selected NAN is modified by updating the pheromone matrix as

$$T_{j,m,\xi}^{S_{NGW}} = T_{j,m,\xi}^{S_{NGW}} + \Delta\tau$$
$$\forall j \in \{1,...,N_{HGW}^{S_{NGW}}\} \quad (14)$$

where the updating pheromone amount is

$$\Delta\tau = (B_{j,m}^{S_{NGW}})^{M \times (\sum_{j=1}^{M} L_{j,m})^{-1}} \quad (15)$$

*HGW Selection:* After selection of NGW, the HGW is selected at the NAN level. In this step, probability of each HGW candidate for the $i = S_{NGW}$, which were constructed at the *NGWs Probabilities* step using (12) is loaded. Then, a HGW is selected for channel m by

$$S_{HGW} = \begin{cases} \arg\max \left( p_{j,m}^{S_{NGW}} \right) & g > G' \\ \text{Roulette Wheel}\left( p_{j,m}^{S_{NGW}} \right) & \text{otherwise} \end{cases} \quad (16)$$

where $S_{HGW}$ is the selected HGW $G'$ is the selection parameter with uniform probability in $0 < G' < 1$. It should be noted that each ant can visit each node for once.

*Local Pheromone Updating:* Pheromone amount of $S_{HGW}$ for the channel $m$ is updated as

$$T_{S_{NGW},m}^{S_{NGW}} = T_{S_{NGW},m}^{S_{NGW}} + \Delta\tau \quad (17)$$

for $j = S_{HGW}$ in (15).

*Global Pheromone Updating:* In order to follow the ACS strategy in pheromone evaporation, the last traditional step of each completed iteration is global pheromone updating as

$$T^i_{j,m,\xi} = \rho T^i_{j,m,\xi}, \quad (18)$$
$$\forall i \in \{1,...,N_{NGW}\}, j \in \{1,...,N^i_{HGW}\}, m \in \{1,...,M\}$$

where the parameter $0 < \rho < 1$ is the evaporation coefficient [6,10,14].

*Final Channel Selection:* Finally, a channel $\omega_j$ is assigned to each HGW based on their corresponding pheromone amount as

$$\omega_j = \arg Max(\frac{1}{\Xi}\sum_{\xi=1}^{\Xi} T_{j \times m \times \xi}) \;. \quad (19)$$
$$\forall m \in \{1,...,M\}$$

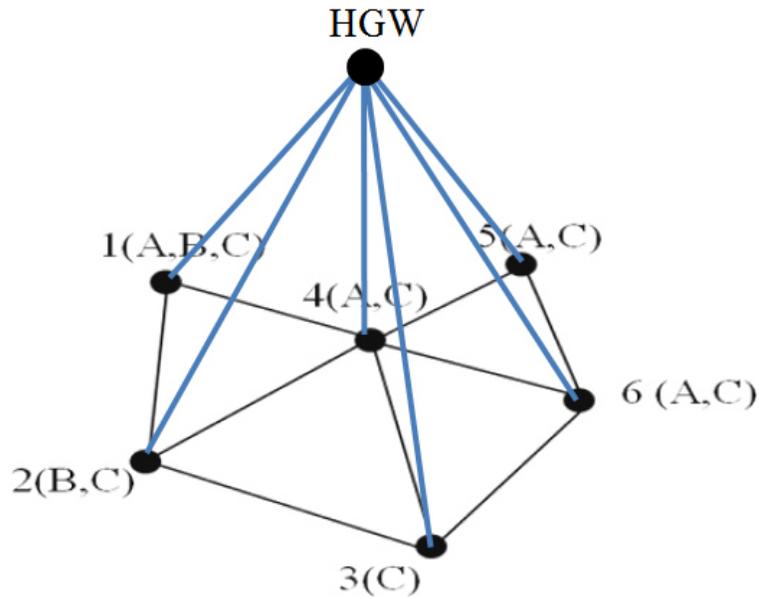

Figure 5 Smart device gate ways (SDGWs) interferences modeled a as a graph problem.

```
Procedure Fuzzy-ACS Cognitive HAN Fair Resource Management Algorithm
  Initialization
  Graph Mapping
  For each Iteration
    For each Channel
      For each Ant
        If any SU exists
          Calculate SDGW Probabilities
          SDGW Selection
          Local Pheromone Updating
        End If any SU exists
      End each Ant
    End each Channel
    Global Pheromone Updating
```

```
End each Iteration
Final Channel Selection
End Procedure Fuzzy-ACS Cognitive HAN Fair Resource Management Algorithm
```

*Figure 6 Proposed Fuzzy-ACS Cognitive home area network (HAN) fair resource allocation algorithm in pseudocode.*

## 4.4. Simulation Results

In this section, performance of the proposed method is analyzed with respect to the utilization functions MSR, MMR, and MPF in comparison with the color sensitive graph coloring (CSGC) method in [15] and a random channel assignment. Then, its stability and convergence is examined for different number of ants as well iterations.

### 4.4.1. Parameters Setting

To do so, a desktop computer with Intel Core2Quad T9300 2.5 GHz CPU and 4 GB of RAM is employed for simulations in MATLAB 2010b. In order to set the parameters of algorithm, different cases have been studied based on trial and error to achieve the best performance. The parameters are set to $C_{max} = 10$, $d_{max} = 4$, $d_{min} = 1$, and $dp = 2$. In simulations, 5 NANs are considered where in which 20 HGWs exist. In order to avoid interference with PUs, each HGW adjusts its transmit power and interference range $ds_{n,m}$ on each channel $m$ by giving the location and channel selection of PUs. Configuration of PUs determines channel availability, reward, and interference constraints, as seen by SUs. Increasing the number of PUs or increasing the protection range, not only can expand the primary protection area but also force affected SUs to reduce their power and thus $ds$. This results in reduction of the available channels as well as channel reward at SUs and therefore degrades spectrum utilization. In addition, the interference among SUs decreases and improves the possibility of spectrum reuse by multiple SUs [15].During simulations, 15 ants are employed and the ACS parameters are considered as $\rho = 0.9$, $G = 0.9$, and $G' = 0.9$.

### 4.4.2. Performance Analyze

In order to analyze performance of the proposed algorithm for different number of available channels, 10 PUs and 20 HGWs in each NAN are considered. As Figure 7 illustrates, increasing the number of channels results in more MSR, MMR and MPF values. The comparison between the Rand, CSGC, and the proposed ACS algorithm shows that the ACS has the best performance among the other approaches under all utilizations. In Figure 8, performance of the proposed algorithm is studied for varying number of PUs for 10 channels and 10 HGWs. As it is illustrated, by increasing the dense of PUs in NANs, the MSR, MMR, and MPF utilizations decrease dramatically. As expected, the random channel assignment has the least performance, while the ACS approach has the best achievement. Increasing the number of HGWs in NAN, results in more interference on available channels. As Figure 9 illustrates, this increment decreases the overall network utilization for all utilization measures. Similar to the previous results, the ACS has the best outcome of utilization while the CSGC performs much better than the random approach. The above results clearly demonstrate that the ACS approach due to the artificial visibility and ants memory can perform much stronger in joint resource management in cognitive HAN of machine to machine networks. However, its stability and convergence will be studied in the following subsection.

Stability of the biologically inspired algorithms is mostly studied by their performance in convergence for varying the number of artificial ants as well as iterations. In Figure 10, the ACS approach is studied with respect to the normalized total cost, in each iteration. In the beginning, the algorithm has many interchanges, however, after 14 cycles it converges.

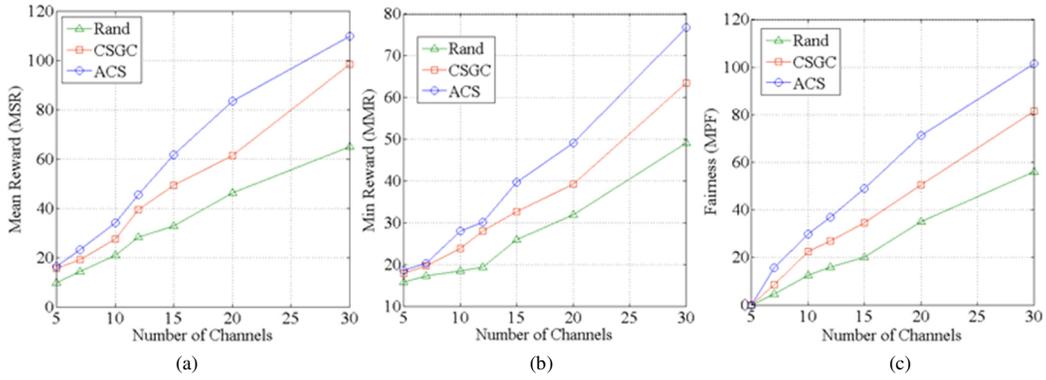

Figure 7 Spectrum assignment with varying number of channels versus three utilization functions: a) MSR; b) MMR; c) MPF.

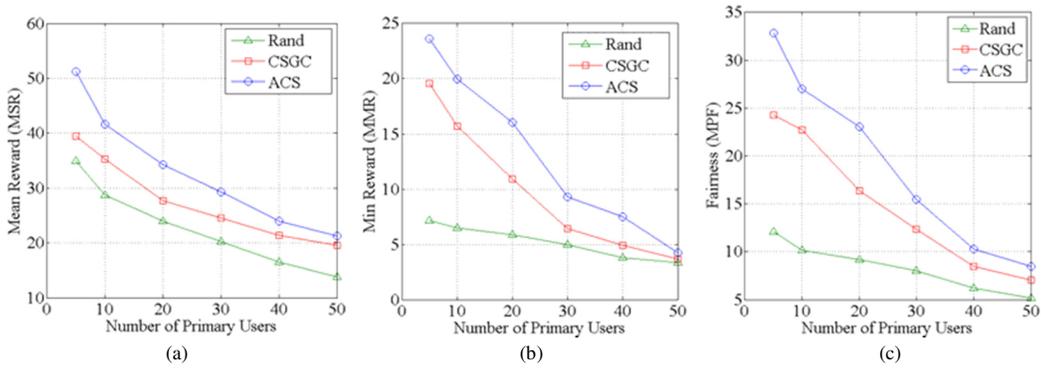

Figure 8 Spectrum assignment with varying number of primary users versus three utilization functions: a) MSR; b) MMR; c) MPF.

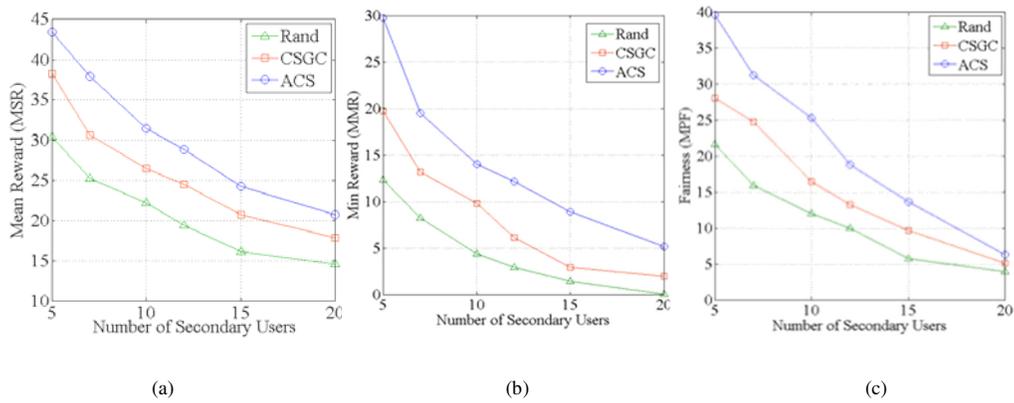

*Figure 9 Spectrum assignment with varying number of secondary users versus three utilization functions: a) MSR; b) MMR; c) MPF.*

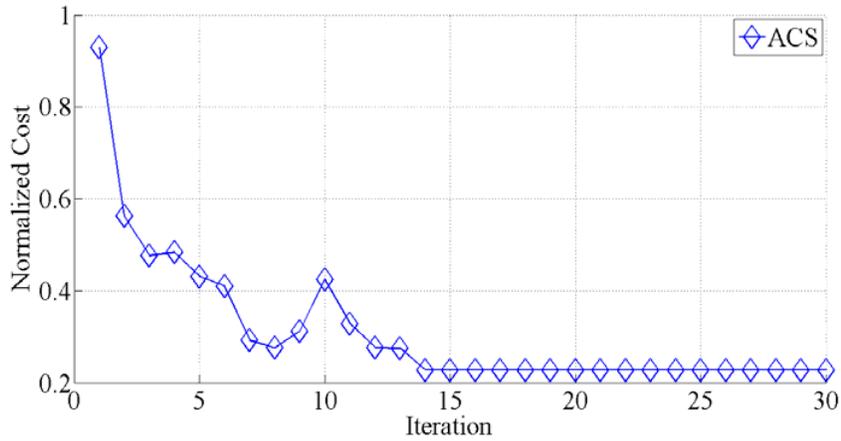

*Figure 10 Performance of the proposed ACS algorithm with respect to normalized cost in each iteration number.*

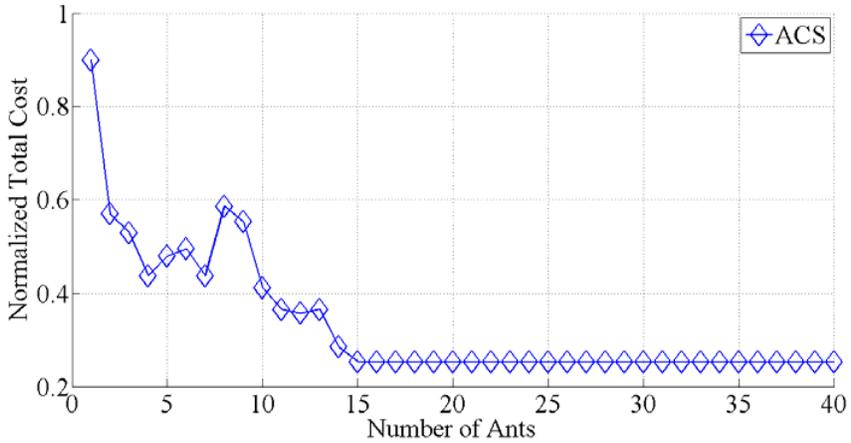

*Figure 11 Performance of the proposed ACS algorithm with respect to normalized total cost for each number of artificial ants.*

Convergence of algorithm in middle iterations is a proof for algorithm stability and proper adjustment [6,9,10,14]. In another study, the algorithm is analyzed for varying number of artificial ants. As Figure 11 demonstrates, the algorithm has converged for 15 numbers of artificial ants. It shows that as the number of ants increases, the system computational load also increases but the normalized total system cost decreases.

The simulations verifies that the proposed ACS is a reliable and stable approach with low computational complexities as well as superior performance for joint resource management in hierarchical cognitive radio based machine to machine communication.

## VI. CONCLUSION AND FURTHER CHALLENGES

In this paper, a review on recent approaches in machine to machine communication is presented. The current methods are advanced in many aspects, however, such methods are not efficient for large-scale machine to machine communication and control. In this report, a new structure for communication between machine on cloud is presented. In this topology, the user has direct control on the utilities, for example in a residential place, and a central optimization and data analytics core control the behavior of machines. The data mining and optimization approaches are utilized to recommend promotional offers to the user. Since in such topology, all the devices are in communication with a control center, many problems such as the lack of communication channels arise. To do deal with this problem as a case study, the hierarchical structure of cognitive radio based communication network in machine to machine communication systems is studied in home area networks (HANs) and corresponding interferences model using graph theory is presented. Then, an ant colony system (ACS) based approach is introduced for fair resource management in the model. Performance study of the proposed procedure has yielded significant results in resource management in cognitive HAN and has prepared a basic study for further developments in the topic.

There are several challenges lie ahead before the cognitive radio based communications networks for fair and high performance resource management in the machine to machine communication networks can be deployed. In order to keep simplicity, the HANs are considered without interference with each other. Also, the users are considered static, while in real scenarios mobile users such as hybrid cars can also be considered as smart grid services. The other issue is QoS-aware policies which should be used in hybrid dynamic spectrum access (HDSA) in NANs and employing cooperative relay techniques into the communications infrastructure for QoS enhancement. Biologically inspired algorithms are state of the art methods for solving NP-hard problems. Performance of such methods in joint resource management in cognitive radio based smart grid networks should be studied. There are also several techniques such as fuzzy system to control performance of such algorithms which all need further study and development by the academia as well as industry.

## References

[1] R. Yu, Y. Zhang, S. Gjessing, C. Yuen, S. Xie, M. Guizani, "Cognitive radio based hierarchical communications infrastructure for smart grid," IEEE Network, vol. 25, Issue 5, pp. 6 – 14, 2011

[2] R. Deng, et. al. "Sensing-Delay Tradeoff for Communication in Cognitive Radio enabled Smart Grid," IEEE International Conference on Smart Grid Communications, pp. 155-160, 2011.

[3] R. Yu, Y. Zhang, Y. Chen, "Hybrid spectrum access in cognitive Neighborhood Area Networks in the smart grid," IEEE Wireless Communications and Networking Conference, pp. 1478 – 1483, 2012.

[4] R. Ranganathan, et. al., "Cognitive Radio for Smart Grid: Theory, Algorithms, and Security," International Journal of Digital Multimedia Broadcasting, vol. 2011, 14 pages, 2011.

[5] I. F. Akyildiz, W. Lee, M. C. Vuran, S. Mohanty "NeXt generation/dynamic spectrum access/cognitive radio wireless networks: A survey," Computer Networks, vol. 50, pp.2127-2159, 2006.

[6] H. Salehinejad, F. Pouladi, S. Talebi, "A Metaheuristic Approach to Spectrum Assignment for Opportunistic Spectrum Access", 17th IEEE International Conference on Telecommunications, pp. 234-238, 2010.

[7] Q. Zhao, B. Sadler "A survey of dynamic spectrum access: Signal processing, networking, and regulatory policy," IEEE Signal Processing magazine: Special Issue on Resource-Constrained Signal Processing, Communications, and Networking, pp.79-89, May, 2007.

[8] S. Haykin, "Cognitive Radio: Brain-Empowered Wireless Communications," IEEE Journal on Selected Areas in Communication, vol. 23, no. 2, pp. 201–20, Feb. 2005.

[9] H. Salehinejad, S. Talebi, "Dynamic Fuzzy Logic-Ant Colony System Based Route Selection System", Applied Computational Intelligence & Soft Computing, Article ID 428270, Vol. 2010.

[10] M. Dorigo, LM. Gambardella, "Ant Colony System: A Cooperating Learning Approach to the

Traveling Salesman Problem", IEEE Trans. Evol. Computing, 1997.
[11] Z. Zhijin, P. Zhen, Z. Shilian, S. Junna, "Cognitive radio spectrum allocation using evolutionary algorithms," Wireless Communications, IEEE Transactions on, Volume: 8 , Issue: 9, pp. 4421 – 4425, 2009.
[12] G. Vidyarthi, A. Ngom, I. Stojmenovic, "A hybrid channel assignment approach using an efficient evolutionary strategy in wireless mobile networks," Vehicular Technology, IEEE Transactions on, vol. 54 , Issue: 5  pp. 1887 – 1895,  2005.
[13] C. Peng, H. Zheng, B.Y. Zhao, "Utilization and fairness in spectrum assignment for opportunistic spectrum access," ACM Mobile Networks and Applications, vol. 11(4), pp. 555–576, 2006.
[14] F. Koroupi, S. Talebi, H. Salehinejad, "Cognitive radio networks spectrum allocation: An ACS perspective," Scientia Iranica, vol. 19, Issue: 3, pp. 767-773, 2012.
[15] Peng, C., Zheng, H. Zhao, B.Y. ''Utilization and fairness in spectrum assignment for opportunistic spectrum access'', ACM Mobile Networks and Applications (MONET), 11(4), pp. 555–576, 2006.
[16] "Population Division World Population to 2300," United Nations Department of Economic and Social Affairs, 2004.
[17] "Networked Society City Index," Ericsson, 2013.
[18] F. Farrahi Moghaddam, "Carbon-profit-aware job scheduling and load balancing in geographically distributed cloud for HPC and web applications," Ph.D. Thesis, ETS, 2014.
[19] "Shaping sustainable cities in the Networked Society," Ericsson, 2011.
[20] "SMART 2020: Enabling the low carbon economy in the information age," The Climate Group, Global e-Sustainability Initiative (GeSI), 2008.
[21] A. Ipakchi, F. Albuyeh, "Grid of the future," IEEE Power and Energy Magazine, vol. 7, no. 2, pp. 52-62, 2009.
[22] "Renewable portfolio standards fact sheet," U.S. Environmental Protection Agency, 2008.
[23] "Regulations," U.S. Environmental Protection Agency.
[24] "What is Telemedicine?," American Telemedicine Association, Washington, D.C., 2011.
[25] H. Salehinejad, S. Rahnamayan, H. R. Tizhoosh, and S. Y. Chen, "Micro-Differential Evolution for High Dimensional Problems," IEEE Congress on Evolutionary Computation, 2014.
[26] H. Salehinejad, S. Rahnamayan, and H. R. Tizhoosh, "Toward Using Type-II Opposition in Optimization: Type-II Opposition-Based Differential Evolution," IEEE Congress on Evolutionary Computation, 2014.
[27] R. V. Zicari, "What are the challenges for modern Data Centers? Interview with David Gorbet,"March 25, 2014
[28] H. Salehinejad, Micro-Differential Evolution: Diversity Enhancement and Comparative Study. Diss. University of Ontario Institute of Technology, 2014.
[29] H. Salehinejad, R. Zadeh, R. Liscano, and S. Rahnamayan, "3D localization in large-scale Wireless Sensor Networks: A micro-differential evolution approach." Personal, Indoor, and Mobile Radio Communication (PIMRC), 2014 IEEE 25th Annual International Symposium on. IEEE, 2014.